\documentclass{ws-p8-50x6-00}

\begin{document}

\title{Symmetry Violation in Heavy Nuclei}

\author{Vladimir Gudkov}

\address{Department of Physics and Astronomy, University of South Carolina, Columbia SC 29208,
USA\\E-mail: gudkov@sc.edu}


\maketitle

\abstracts{ The advantage of searching for violation of Time
Reversal Invariance in neutron induced reactions using relative
measurements of Time and Parity violating effects is discussed.
This approach gives the enhancement of $T$-violating effects by
many orders of magnitude and, at the same time, decrees
theoretical uncertainties and experimental errors. }

\section{Introduction}
For a long time heavy nuclei were considered for the study of
symmetry violations mostly due to enhancement factors related to
the complex nuclear structure. Classical examples are experiments
that measure parity violating correlations in nuclear decays and
in nuclear reactions. It is well known that due to nuclear
enhancements\cite{SF,BG1} the parity violating effects in nuclei
could be as large as $\sim 10^{-1}$ in comparison to its
"natural" scale $\sim 10^{-7}$ for the simplest few nucleon
systems. This feature of heavy nuclei has been explored for the
gathering of unique information about statistical properties of
nuclei and for further development of the understanding of
nuclear structure (see, for example ref.\cite{Al,YMex,JDB,GaryM}).
However, it is practically impossible to obtain information about
weak interactions using these precisely measured nuclear
$P$-violating effects.  The main reason for that is the extremely
complicated structure of nuclear wave functions of the excited
states of heavy nuclei.  Therefore, $P$-violating effects in
heavy nuclei could be treated  by using mainly statistical
approaches.

Time Reversal Invariance ($TRI$) is another example of a
fundamental symmetry that could be enhanced in heavy nuclei.
Since the origin of the possible $TRI$ violation is not yet
understood, and even a proof of the $TRI$ violation
($T$-violation) does not exist, it is very important to search
for $T$-violation using all opportunities. To avoid a possible
misunderstanding, we clarify the meaning of the proof of
$T$-violation.

From the point of view of a local Lorentz invariant field theory
with ordinary relations between spin and statistics,
$T$-violation can be related through the $CPT$ theorem to
$CP$-violation which was discovered experimentally in the
$K^0$-meson decays.  This kind of $T$-violating interaction
violates parity, too. In this framework there is another
possibility for $T$-violation due to $T$-violating $P$-conserving
($C$-odd and $P$-even) interactions (see, for example
ref.\cite{VG1,RM}). However, an extension of the theory beyond
the locality of the interactions or with violation of Lorentz
invariance and/or the spin-statistics relations might violate the
$CPT$ theorem. Therefore, the
 $CPLEAR$ experimental result\cite{CL} could be considered as
evidence of $T$-violation in $K^0$-meson decays under the
assumption of the $CPT$ invariance.

It should be noted that in $K^0$-meson decays the $CP$-violating
interactions belong to the $\Delta S = 1$ sector. However, some
models (for example, Minimal Super Symmetry Model) predict large
$CP$-violating effects for the $\Delta S = 0$ sector and a
suppression of $CP$ violation with a change of strangeness
$\Delta S = 1$. From this point of view a low energy physics
($\Delta S = 0$) is very interesting because it can give
independent additional information about $CP$ violation.

\section{$T$-violation and neutron reactions}

Taking into account the above arguments one can see the
importance of the search for $T$-violation in low energy physics.
It equally applies to both nuclear reactions and measurements of
neutron and atomic dipole moments. However, there are some
specific features, both theoretical and experimental, which
distinguish neutron induced reactions from a wide class of low
energy experiments. Let us consider some of them.

  The important
questions to address in the search for the $T$-violation effects
are: how well can the effect be calculated, and how well do we
know existing restrictions on the $T$-violating coupling
constants? These questions are  very important for the
interpretation of a positive experimental result. However, they
are much more important for estimating the upper limit on the
unknown $T$-violating constants in case the experiment will give
only an upper bound on the value of the effect.  In that relation,
one recalls a history of calculations of the electric dipole
moment ($EDM$) for a simple system like the neutron. We can see
that even for rather well defined mechanisms of $CP$ violation,
such as the standard Kobayashi-Maskawa model and Weinberg model of
spontaneous $CP$-violation in Higgs sector, the precisely
estimated value had changed over time by many orders of magnitude.
The reason for that is a discovery from time to time of different
contributions from strong interactions (such a strange quark
chromoelectric dipole moment or multi gluon $CP$-odd operator)
which change drastically the estimated values of the neutron
$EDM$. The situation could be much worse for more complicated
systems. One of the recent examples of possibly similar
difficulties is the discrepancy (by about $2.5 \sigma$) between
the most precise measurement\cite{CsEx} of $P$ violation in the
atom of $^{133}Cs$ and its accurate theoretical calculations. This
situation rises questions about a possible discovery of "new
physics"  or, probably, about simply an overestimation of the
calculated accuracy.

Is it possible to avoid such difficulties in the search for an
unknown phenomena like $T$ violation?  Fortunately, for some
cases of neutron induced reactions the answer is {\it yes}. The
reasons for that are structural similarities between interactions
with different symmetries and the possibility to measure
different effects corresponding to a violation of these symmetries
simultaneously. To illustrate the general idea, let us consider
different types of interactions in the matrix momentum
representation. Then, $P$- and $T$-conserving strong interactions
can be represented by an orthogonal (real symmetric) matrix:

\begin{displaymath}
\mathbf{V_{str}}= \left( \begin{array}{ccccc}
  v_{00} & 0       & \ldots  & v_{ik} & \ldots \\
  0      & v_{11} & \ldots & \ldots  & \ldots \\
  \vdots & \vdots & \ddots & \ldots  & \ldots \\
  \vdots & v_{ki} & \vdots & v_{ii}  & \ldots \\
  \vdots & \vdots & \vdots & \vdots  & \ddots
\end{array} \right),
\end{displaymath}
with $v_{ik} = 0$ for angular momenta $i$ and $k$ with different
parities.

In the same representation, parity violating interactions look
like
\begin{displaymath}
\mathbf{V_{P}}= \left( \begin{array}{ccccc}
  0       & v_{01}& \ldots  & v_{ik} & \ldots \\
  v_{10} & 0      & \ldots & \ldots  & \ldots \\
  \vdots & \vdots & \ddots & \ldots  & \ldots \\
  \vdots & v_{ki} & \vdots & 0       & \ldots \\
  \vdots & \vdots & \vdots & \vdots  & \ddots
\end{array} \right),
\end{displaymath}
with $v_{ik} = 0$ for angular momenta $i$ and $k$ with the same
parities. In this representation $T$-violating interactions bring
anti-symmetric imaginary parts to all matrix elements. Therefore,
$T$-violating $P$-conserving interactions are
\begin{displaymath}
\mathbf{V_{T}}= \left( \begin{array}{ccccc}
  0      & 0       & \ldots  & +{\it i}w_{ik} & \ldots \\
  0      & 0       & \ldots & \ldots  & \ldots \\
  \vdots & \vdots & \ddots & \ldots  & \ldots \\
  \vdots & -{\it i}w_{ki} & \vdots & 0        & \ldots \\
  \vdots & \vdots & \vdots & \vdots  & \ddots
\end{array} \right),
\end{displaymath}
and $T$-violating $P$-violating interactions are
\begin{displaymath}
\mathbf{V_{PT}}= \left( \begin{array}{ccccc}
  0       & +{\it i}w_{01}& \ldots  & +{\it i}w_{ik} & \ldots \\
  -{\it i}w_{10} & 0      & \ldots & \ldots  & \ldots \\
  \vdots & \vdots & \ddots & \ldots  & \ldots \\
  \vdots & -{\it i}w_{ki} & \vdots & 0       & \ldots \\
  \vdots & \vdots & \vdots & \vdots  & \ddots
\end{array} \right).
\end{displaymath}
One can see the similarity in the structure of the interactions
with the same parity properties and different Time Reversal
symmetries: $T$-violation brings only an asymmetric
 phase. However, the parity violation changes structure of the
 interactions dramatically.  Therefore, we can
 avoid the influence of nuclear structure on the measurable parameter
  by choosing it as an appropriate ratio of $T$-violating and $T$-conserving
  effects.  The example for one of the possible choices will be considered
  in next section.

 Prior to that, let us recall a general experimental advantage of
 neutron reactions -- extremely high energy resolution. This
 property gives us the opportunity to resolve very narrow resonances and to
 profit from the resonance enhancement factor in the search for $T$ violation.
 The typical
 neutron energy for the resonance reactions is about $1eV -
 100eV$. When neutron is captured by nucleus, the compound nuclear
 excitation energy is about $6MeV$. Then, without any effort the
 energy resolution is about $2\cdot 10^{-7}$. It could be even much
 better in real experiments (by about $\sim 10^{-11}$). Taking into account
 the high experimentally achieved level of neutron polarization
 and the high flux with good time structure at the Spallation Neutron
 Source, one can conclude that neutron reactions are very
 promising for precise measurements of $T$ violation.

\section{$P$- and $T$- violating effects in neutron scattering}

To illustrate the advantage of the search for $T$-violation in
heavy nuclei with neutrons, we consider the case of simultaneous
Parity and Time Reversal ($PT$) violation. One of the effects
related to $PT$ violation is the difference of total cross
sections $\Delta \sigma_{PT}$ in the transmission of polarized
neutrons through a polarized target for opposite neutron spin
orientations. (For some other mechanisms of $T$ violation and
other effects, see the paper\cite{VG1} and references therein.
The experimental proposal to search for $PT$ violation in a
different geometry is discussed in the paper.\cite{YM}) This
effect is proportional to the $T$-odd correlation
$(\vec{\sigma}\cdot [\vec{k}\times \vec{I}])$ between spin
$\vec{\sigma}$ and momentum $\vec{k}$ of neutron and nuclear spin
$\vec{I}$. Let us also consider the corresponding $P$-violating
$T$-conserving difference of total cross sections $\Delta
\sigma_{P}$ in the transmission of polarized neutrons through
unpolarized target which is proportional to the correlation
$(\vec{\sigma}\cdot \vec{k})$. Using the optical theorem, one can
represent these differences of total cross sections in terms of
differences of zero angle scattering elastic amplitudes for
opposite neutron spin orientations along axis $[\vec{k}\times
\vec{I}]$ for $PT$-odd effect and along $\vec{k}$ for $P$-odd
effect :
\begin{equation}
 \Delta \sigma_{PT} = {{4\pi}\over k}{\it Im}(f_{\uparrow} - f_{\downarrow})
\label{eq:dpt}
\end{equation}
and

\begin{equation}
 \Delta \sigma_P = {{4\pi}\over k}{\it Im}(f_- - f_+).
 \label{eq:dp}
\end{equation}

One can calculate both these parameters using distorted wave Born
approximation in the first power of parity and time reversal
violating interactions (see, for example ref.\cite{BG1}).  Then
the symmetry violating amplitudes can be written as
\begin{equation}
t^{fi}_{P,PT} = <{\Psi^-_f}|V_{P,PT}|{\Psi^+_i}>,
\end{equation}

 where $\Psi^{\pm}_{i,f}$ are the eigenfunctions of the nuclear T-invariant
 Hamiltonian with  the  appropriate boundary conditions:
\begin{equation}
 \Psi^{\pm}_{i,f}=\sum_k a^\pm_{k(i,f)}(E)\; \phi_k + \sum_m\int
b^{\pm}_{m(i,f)}(E,E')\; \chi^{\pm}_m(E')\; dE'.
 \label{eq:wf}
\end{equation}
Here $\phi_k$ is the wave function of  the $k^{th}$
 compound-resonance and $\chi^{\pm}_m(E)$ is the potential
 scattering wave function in the channel $m$.
The coefficient
\begin{equation}
 a^\pm_{k(i,f)}(E)={\exp{(\pm i\delta_{i,f})}\over {(2\pi)^{1\over 2}}}{{(\Gamma^{i,f}_k)^{1\over 2}}\over {E-E_k\pm{i\over
   2}\Gamma_k}}
\end{equation}
describes compound nuclear resonances reactions and the
coefficient $b^{\pm}_{m(i,f)}(E,E')$ describes potential
scattering and interactions between the continuous spectrum and
compound resonances. (Here $E_k$, $\Gamma_k$, and $\Gamma^i_k$ are
the energy, the total width, and the partial width in the channel
$i$ of the $k$-th nuclear compound resonance, $E$ is the neutron
energy, and $\delta_i$ is the potential scattering phase in the
channel $i$; $(\Gamma^i_k)^{1\over 2} = (2\pi )^{1\over 2}
 <{\chi_i(E)}|V|{\phi_k}>$,
 where $V$ is a residual interaction operator.)

Therefore, in general, there are many mechanisms that violate the
symmetries in nuclei and theoretical descriptions of symmetry
violating effects could be rather complicated. However, it was
shown\cite{BG1} (and confirmed by many experiments) that for the
bulk of heavy nuclei the dominant mechanism for parity (and $PT$)
violation is the mechanism of symmetry mixing on the compound
nuclear stage. This mechanism could be described by the first
term in Eq.~(\ref{eq:wf}). For our illustrative purposes we
consider the simplest case of a two resonance approximation. This
approximation is reasonably good for many heavy nuclei in the low
neutron energy  region $E \sim 1 eV - 10 eV$, since the
characteristic energy difference between compound neutron
resonances with the same spin and parity is usually about $10 eV
- 100 eV$. Assuming that main effects could be described by
mixing of the nearest $s$-wave and $p$-wave resonances, one
derives the symmetry violating amplitudes as:

\begin{equation}
<{p}|t|{s}> = - {1\over{2\pi}}
{{(v+iw)(\Gamma^n_s\Gamma^f_p)^{1\over
2}}\over{(E-E_s+i\Gamma_s/2)(E-E_p+i\Gamma_p/2)}}{\it
e}^{i(\delta^n_s + \delta^n_p)}, \label{eq:ps}
\end{equation}
and
\begin{equation}
<{s}|t|{p}> = - {1\over{2\pi}}
{{(v-iw)(\Gamma^n_p\Gamma^n_s)^{1\over
2}}\over{(E-E_s+i\Gamma_s/2)(E-E_p+i\Gamma_p/2)}}{\it
e}^{i(\delta^n_p + \delta^n_s)}.
 \label{eq:sp}
\end{equation}

The matrix elements $v$ and $w$ are real and correspond to the
real and imaginary parts of the mixing  matrix element between
$s$- and $p$-wave compound resonances for the sum of $P$- and
$PT$-violating operators $V_{P}$ and $V_{PT}$
\begin{equation}
 v+iw = <{\phi_p}|V_{P}+V_{PT}|{\phi_s}>.
  \label{eq:me}
\end{equation}
 The matrix element $v$ violates parity but preserves Time Reversal
Invariance. The matrix element $w$ violates both $P$- and
$T$-invariance. It contributes to the total mixing matrix element
Eq.~(\ref{eq:me}) as an imaginary anti-symmetric part which
results in its opposite sign for the amplitudes in
Eq.~(\ref{eq:ps}) and Eq.~(\ref{eq:sp}).

It is important that these amplitudes describe both $P$-violating
and $PT$-violating processes.  Moreover, matrix elements $v$ and
$w$ in these amplitudes are matrix elements calculated using
exactly the same wave functions.

Another remarkable fact is that the difference of amplitudes $(f_-
- f_+)$ for $P$-violating effect in Eq.~(\ref{eq:dp}) is
proportional to the sum of the symmetry violating amplitudes
(Eq.~(\ref{eq:ps}) and Eq.~(\ref{eq:sp})) but the difference of
amplitudes $(f_{\uparrow} - f_{\downarrow})$ for $PT$-violating
effect in Eq.~(\ref{eq:dpt}) is proportional to the difference of
the same amplitudes (Eq.~(\ref{eq:ps}) and Eq.~(\ref{eq:sp})).
This results in the same energy dependencies for both $P$- and
$PT$-violating effects. Indeed, taking into account all numerical
factors one gets:
\begin{equation}
 \Delta\sigma_{PT} = - {{2\pi G_J}\over
k^2}{{w(\Gamma^n_s\Gamma^n_p(S))^{1\over 2}}\over{[s][p]}}[(E -
E_s)\Gamma_p + (E - E_p)\Gamma_s],
\end{equation}

and
\begin{equation}
 \Delta\sigma_{P} = {{2\pi G_0}\over
k^2}{{w(\Gamma^n_s\Gamma^n_p)^{1\over 2}}\over{[s][p]}}[(E -
E_s)\Gamma_p + (E - E_p)\Gamma_s],
\end{equation}
where  $[s,p]=(E-E_{s,p})^2+{{\Gamma^2_{s,p}}/4}$, $G_J$ and $G_0$
are spin factors; $J$ and $S$ are compound nuclei and channel
spins (see details in ref.\cite{BG1,BG2,VG1}). Due to the
similarity in these two parameters it is obvious that the
$PT$-violating effect has the same resonance enhancement as the
$P$-violating one. Also, they have similar dynamic enhancement
factors. Therefore, one can conclude that the $PT$-violating
effect has about the same nuclear enhancement factors ($\sim
10^6$) as the $P$-violating one (see, for details ref.\cite{BG2}).

Taking the ratio of $\Delta\sigma_{PT}$ and $\Delta\sigma_{P}$
parameters from the same experiment at the same neutron energy one
can extract the ratio of $PT$- and $P$-violating matrix elements
$<\lambda > = {w/v}$ eliminating nuclear reaction uncertainties
and experimental uncertainties related to the absolute
normalization:
\begin{equation}
 {{\Delta\sigma_{CP}}\over{\Delta\sigma_{P}}} = \kappa
 (J){w\over{v}},
\end{equation}
where $\kappa (J)$ is the calculable spin dependent
coefficient.\cite{VG2}

Now it is worthwhile to recall that the extracted ratio $<\lambda
> = {w/v}$ is not the ratio of independent matrix elements
of complicated compound nuclear states, but rather the ratio of
matrix elements with exactly the same wave functions and very
similar operators.\cite{VG2} This ratio might be calculated for
each particular nucleus with quite good accuracy. In the simplest
case one can do it using one particle $P$-violating and
$PT$-violating potentials
\begin{equation}
V_P = {G\over{8^{1/2}M}} \{ ({\vec{ \sigma}\cdot \vec{p}}),\rho
(\vec{r})\} _+,
\end{equation}

\begin{equation}
 V_{PT} = {{iG\lambda }\over{8^{1/2}M}} \{ ({\vec{\sigma}\cdot
\vec{p}}),\rho (\vec{r})\}_-
\end{equation}
 where $G$ is the weak
interaction Fermi constant, $M$ is the proton mass, $\rho (\vec{
r})$ is the nucleon density, $\vec{p}$ is the momentum of the
valence nucleon, and $\lambda = {g_{PT}/{g_{P}}}$ is the ratio of
$PT$-violating to $P$-violating nucleon - nucleon coupling
constants. Then one obtains
\begin{equation}
<\lambda > = {\lambda\over{1+2\xi}},
\end{equation}
 where $\xi \sim (1 - 7)$ (for detailed discussions see papers\cite{VG2,TH,Kh,DN}).

 Now we come to the result that the simultaneous measurement of $PT$-
 and $P$-violating effects affords the opportunity to extract the
 ratio of $PT$- and $P$-violating nucleon coupling constants,
 $\lambda = {g_{PT}/{g_{P}}}$, eliminating nuclear uncertainties
 and some experimental uncertainties.

 Let us compare the accuracy for the parameter $\lambda_{exp} \sim 10^{-4}$
 which could be achieved\cite{YMp} at the Spallation Neutron
 Source with theoretical expectations for the ratio
 $\lambda = {g_{PT}/{g_{P}}}$. To do this we assume that $PT$ violation is
 related to the $CP$ violation under the $CPT$-theorem. (All
 existing calculations for the parameter $\lambda$ have been done for the $CPT$-invariant
 models.)
 The estimated values\cite{VG2,H1,H2,VG3} of the parameter $\lambda$ for
 some models of $CP$ violation are given in
 Table~\ref{tab:lam}. The last two rows in Table~\ref{tab:lam}
 correspond to limits on the parameter
 $\lambda$ obtained from the experimental results of measurements of neutron
  and atomic electric
 dipole moments. The parameter $\lambda$ of neutron electric dipole
 moment\cite{NEDM} (EDM) has been
  calculated using phenomenological  $\pi$-meson one loop mechanism. The
  parameter $\lambda$ of atomic EDM ($^{199}Hg$) \cite{AEDM} is
  addopted from the paper.\cite{AtL}

\begin{table}[t]
\caption{The relative values $\lambda$ of the $CP$-violating
nucleon coupling constants.\label{tab:lam}}
\begin{center}
\footnotesize
\begin{tabular}{|c|l|}
\hline {Model}  &{$\lambda$}\\
\cline{1-2} Kobayashi-Maskawa &  $\leq 10^{-10}$ \\
 Right-Left model & $\leq 4\times 10^{-3}$ \\
 Horizontal symmetry & $\leq  10^{-5}$ \\
Weinberg model (charged Higgs bosons) & $\leq 2\times 10^{-6}$ \\
Weinberg model (neutral Higgs bosons) & $\leq 3\times 10^{-4}$ \\
$\theta$-term in $QCD$ & $\leq 5\times 10^{-5}$ \\
Neutron EDM (one $\pi$-loop mechanism) & $\leq 4\times 10^{-3}$ \\
Atomic EDM ($^{199}Hg$) & $\leq 2\times 10^{-3}$ \\

\hline
\end{tabular}
\end{center}
\end{table}

It should be noted that due to model uncertainties and low energy
$QCD$ corrections, the reasonable accuracy of the above
estimations is usually about one order of magnitude. Therefore,
the comparison of the $\lambda_{exp}$ and $\lambda$ presented in
  Table~\ref{tab:lam} leads to the observation that the possible
  experiment for the search
for $T$ violation at the SNS has reasonably good sensitivity to
test many models of $CP$ violation.

For the completeness of our discussion, it is necessarily to
mention that a real experimental setup for these types of
experiments should be quite complicated in order to eliminate the
Final State Interactions (FSI) and to control them during the
experiment. It is well known that a $T$-odd correlations have no
relations to the Time Reversal Invariance (TRI) in general. This
is different from the case of Parity Invariance. The reason for
this difference is the fact that the Time Reversal operator,
unlike the Parity operator, has no eigenstates nor eigenvalues.
Therefore, it leads not to restrictions on the reaction amplitude
but to a relation between amplitudes for two different processes.

However, under special circumstances TRI could be related to
$T$-odd correlations with some restrictions. The well known
example is the $T$-odd correlations in nuclear (particle) decays
when the decay can be described in the first Born approximation.
In that case the unitarity condition of the scattering matrix
leads to its hermicity, which is an additional condition on the
amplitude resulting in the similarity between $T$-odd and $P$-odd
operators and, as a consequence, in the connection of $T$-odd
correlations to the $T$-violating interactions. When the other
Born terms of higher order (usually called the FSI) become
significant, the hermicity condition breaks down and the $T$-odd
correlations could be produced without $T$-violating interactions.

In our case the neutron transmission is an elastic scattering at
the zero angle. Then, the initial and final states in this process
coincide. It gives us the additional condition on the scattering
amplitude without any reference to the intensity of the
interactions.  Therefore, $T$-odd correlations in the elastic
scattering always are related to $T$-violating interactions.

In the experiment there is a possibility of neutron depolarization
due to the neutron spin precession in magnetic fields, strong
spin-spin interactions, and weak interactions. All these effects
could destroy the elasticity of the process. Fortunately, it has
been shown that these interference effects can be decreased or
compensated with a given accuracy, and after that, could be kept
under control during the experiment.
 For detailed discussions of these problems see papers\cite{YM,VG1,BG3,VG3,St,Kab} and
references therein.

\section{Summary}

To summarize, the considered advantages of the search for
$T$-violating effects in neutron reactions on heavy nuclei at the
SNS, we emphases again that the unambiguous experimental results
are guaranteed by the comparing the proper symmetry violating
effects in the same experimental setup. It does not only give us
the possibility to decrease the experimental uncertainties, but,
what is much more important, it allows us to measure relative
parameters which are almost free from usually unavoidable
theoretical uncertainties. This feature, in the combination with
the large nuclear enhancement ($\sim 10^6$), and the possibility
to get rid of final state interactions at any desirable level,
gives us a unique opportunity to search for $T$ violation at the
SNS. The $T$-violating effects in neutron reactions are a pure
test of the $T$ violation with no reference to the $CPT$ theorem.
It is also important because low energy processes test $\Delta S
= 0$ sector of the $CP$ interactions which is out of reach for
high energy experiments.

\end{document}